\documentclass{eas}
\usepackage{graphicx}
\begin{document}

\title{Directional detection of Dark Matter} 
\author{F. Mayet}\address{Laboratoire de Physique Subatomique et de Cosmologie, Universit\'e Joseph Fourier Grenoble 1,
  CNRS/IN2P3, Institut Polytechnique de Grenoble, Grenoble, France}
\author{J. Billard}\sameaddress{1}
\author{D. Santos}\sameaddress{1}
%
%
\begin{abstract}
Directional detection is a promising Dark Matter search strategy.  
Taking advantage on the rotation of the Solar system around the galactic center through the Dark Matter halo, 
it allows to show a direction dependence of WIMP events. It requires  the simultaneous measurement of the energy  and the 3D track 
of low energy recoils,  which is a common challenge for all current projects of directional detectors. 
The third CYGNUS workshop on directional dark matter  detection has brought together the scientific community 
working on both theoretical and experimental aspects of the subject. 
In this paper, we give an introductory revue of directional detection of Dark Matter, focusing on the main recent progresses.
\end{abstract}
\maketitle

\section{Introduction}
Directional detection of Dark Matter has been first proposed  as {\em a powerful tool}
to identify genuine WIMP events as such, {\em even with a low angular resolution detector}  (\cite{spergel}). More than twenty years later,
we give a state of the art revue of directional detection of Dark Matter. Two points will be addressed~: 
\begin{itemize}
\item can directional detection bring something new to the field of Dark Matter search ? 
This is obviously a major issue,  given the timescale to build a large directional TPC. 
\item  what are the main key experimental issues that must be addressed in order to access such promising results ?
\end{itemize}

%
%
\section{Directional detection}
\subsection{Directional detectors}
Following early experimental works~(\cite{gerbier.direction,buckland}), several Dark Matter directional detectors (\cite{white}) are being 
developed and/or operated : DM-TPC~(\cite{dmtpc}), DRIFT~(\cite{drift}),  D$^3$~(\cite{d3}), Emulsions~(\cite{emulsions}), 
MIMAC~(\cite{mimac}) and NEWAGE (\cite{newage}). 
Directional detection   requires  the simultaneous measurement of the recoil energy ($E_R$) and the 3D track ($\Omega_R$) of low energy recoils, 
thus  allowing to evaluate the double-differential spectrum $\mathrm{d}^2R/\mathrm{d}E_R\mathrm{d}\Omega_R$ down to the energy threshold.
This can be achieved with low pressure gaseous detectors (TPC) and several gases have been suggested : 
$\rm CF_4$, $\rm ^{3}He$, $\rm C_4H_{10}$ or $\rm CS_2$.

\subsection{Experimental issues}
There is a worldwide effort toward the development of a large TPC devoted to directional detection (\cite{white}) and all current projects
face common challenges amongst which the reconstruction of low energy tracks seems to be the main one. 
In the following, we discuss the key experimental issues for directional detection.

\subsubsection{Track reconstruction} 
As far as directional detection is concerned, the estimation of the initial recoil direction is compulsory.  This gives an intrinsic limitation of this detection strategy as recoil  tracks in low
 pressure gaseous detectors would encounter a rather large angular dispersion ("straggling" effect). Then, when measuring tracks in a 
 gaseous TPC, the electron drift properties implies  a transverse and longitudinal diffusion which  contributes 
 to the angular resolution.\\ 
 Hence,  data of upcoming directional detectors should suffer from rather large angular resolution. 
 Dedicated data analysis is needed 
(\cite{Billard.cygnus}) and experimental evaluation of the angular resolution should be done through detector commissioning, 
using e.g. an ion beam or neutron field. A degradation of the angular resolution results in 
a WIMP-induced distribution  getting less anisotropic and hence closer to the  expected background one.\\
The track spatial resolution is also an issue worth being mentionned. It includes resolution in the anode plane as 
well as along the third dimension (drift space). As shown in (\cite{Billard.cygnus}), a good spatial resolution, $\mathcal{O}(mm)$,  
could be obtained in principle, thus opening the way  to detector fiducialization to reject surface events.\\
Other track observables, such as the track length or differential angular deviation, may be used to discriminate electrons from recoils.

\subsubsection{Sense recognition}
Not only should the track be
3D-reconstructed, but its sense should also be retrieved from the data analysis. 
Without sense recognition, the expected WIMP-induced distribution becomes less anisotropic 
and thus gets closer to the expected background event distribution. This induces an obvious loss of discrimination power.\\ 
As outlined in (\cite{Billard.cygnus}), an asymmetry between upgoing and downgoing tracks is expected, due to 
two different effects. First, the angular dispersion of recoiling tracks should result in a shape asymmetry as 
the beginning of the track should be more rectilinear than its end. Second, a charge collection asymmetry is expected as  $dE/dx$ is 
decreasing with energy at low recoil energy. Hence, more primary electrons should be generated at the beginning of the track.\\
Even though several experimental progresses have been done (\cite{dujmic,burgos,majewski}), sense recognition 
remains a key and challenging experimental issue for directional detection of Dark Matter. In particular, it 
should still be demonstrated that sense recognition may be achieved at low recoil energy, where most WIMP events reside, 
and with which efficiency. For a given directional detector, we argue that the main concern on the head-tail subject is :  {\em how much sense recongnition can be
achieved ?} Indeed,   directional data should be only partialy sense-recognized, {\it i.e.} a strong dependence 
of the sense recognition efficiency  is expected on the energy and the drift distance.

\subsubsection{Energy threshold} 
As for direction-insensitive direct detection, the energy threshold plays a key role for directional detection. 
It is worth emphasizing that it is the lowest energy at which both the initial recoil direction and the energy can be retrieved, what makes it
even more challenging for directional detection. In particular, this directional threshold is higher than the threshold 
for simply detecting recoils. Indeed, a low energy recoil (a few keV) in a low pressure TPC presents a short track length, implying  a small number of images, and a 
   large angular dispersion, implying a  loss of the direction information. The directional energy threshold is closely related 
   to the gas pressure, the target choice,   as well as  the read-out and data analysis
strategy.  There are two main and competing consequences when increasing the energy threshold : a reduction of the number of the  
expected WIMP events  and a sensitivity to the most anisotropic part of the WIMP induced recoil distribution.

\subsubsection{Residual background contamination} 
Zero background is often referred to as the ultimate 
goal for the next generation of direct detection experiments in 
deep underground laboratories. However, owing to the large intrinsic difference between the WIMP-induced and
background-induced  spectra, directional detection could accommodate to a sizeable background 
contamination (sec.~\ref{sec:exclu}). It suggests the idea that a light shielding might be sufficient, thus allowing to reduce  
muon-induced neutron background (\cite{Mei:2005gm}).\\ 
Discrimination of background electron recoils
from nuclear recoils remains of course a fundamental requirement of experiments aiming to detect WIMP dark matter. For a gaseous
directional detector, this could be achieved by means of a selection on the energy/track-length, as for a given energy an electron 
track is expected to be much longer than a recoil one. However, it should still be demonstrated which rejection power 
can be obtained with such a selection.\\
Nevertheless, the background rate estimation remains also as a 
key point of the directional data analysis strategy, not for the value itself but for the fact that  a wrong
background estimation may induce bias for Dark Matter parameters.

\subsubsection{Energy measurement} 
The difficulties encountered by the various directional projects in energy measurement are not specific to the directional strategy 
but to the choice of a low pressure gaseous TPC and the need to measure low energy recoils. First, 
a precise energy measurement requires a precise calibration and hence low energy references should be used. Second, the detector allows to
measure the ionization energy which should then be converted to a recoil energy thanks to the knowledge of 
the ionization quenching factor. For a given gas mixture, this quantity needs to be precisely measured  (\cite{santos.q}).

\subsection{Directional target}
As outlined above, the reconstruction of low energy tracks is the main challenge for the future of directional detection. 
It follows that the target nucleus must be light to maximize the track length and, in the case of gaseous detectors, the pressure
must be as low as possible, leading to rather small detector masses as the volume cannot be arbitrarly large. One may then come to the
conclusion that directional detection strategy should focus on spin-dependent  interaction to be competitive with planned and existing
direct detectors. The detector design  should be flexible enough to be able to run with different targets.\\
Then, the ideal directional target is a light nucleus with non-vanishing spin. Leading candidates include : $\rm ^{1}H$, $\rm ^{3}He$ and 
$\rm ^{19}F$  which has been early suggested as
a golden target for spin-dependent  dark matter searches (\cite{ellis.spin}).  Thanks to its 
good ionization  properties (\cite{caldwell}), $\rm CF_4$ is planned  as a sensitive medium for 
most upcoming directional detectors (\cite{white}).\\  
In the following,  we present the case for a low exposure (30 kg.year)  $\rm CF_4$ TPC, operated at 
low pressure and  allowing 
3D track reconstruction, with sense recognition down to the energy threshold.

\section{Directional detection : a powerful tool ?}
 Directional detection strategy  consists in searching for a forward/backward asymmetry in the 
 distribution of WIMP events with respect to the direction of motion of the Solar system, which  
 happens to be roughly in the direction of the Cygnus constellation ($\ell_\odot = 90^\circ,  b_\odot =  0^\circ$ 
 in galactic coordinates).  As the background distribution is expected to be isotropic in the galactic restframe, one expects a 
 clear and unambiguous difference between the WIMP signal and the background one.

\begin{figure}[t]
\begin{center}
  \includegraphics[scale=0.47]{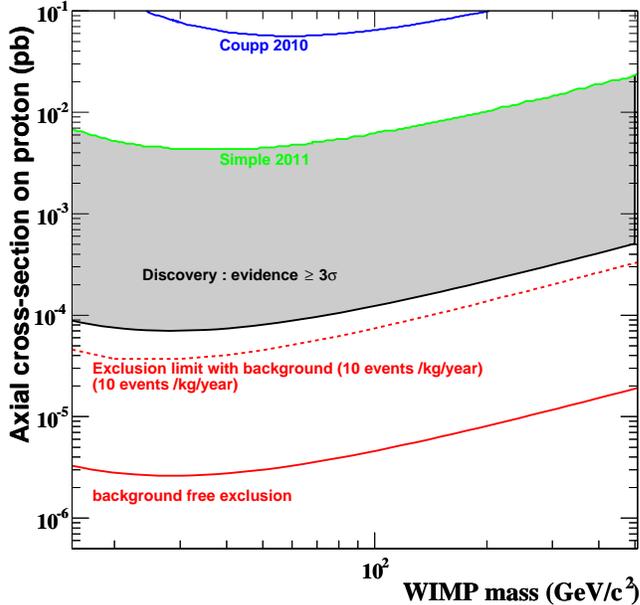}
\caption{\it  
Spin dependent cross section on proton (pb) as a function of the WIMP mass ($\rm GeV/c^2$), in the     pure-proton approximation.  
Exclusion limits from (\cite{coupp}) (dark blue) and  (\cite{simple}) (dark green) are shown. The projected exclusion limit of a forthcoming 
directional detector (30 kg.year) is presented in two cases : 
background-free (light blue solid line) and with a background rate of 10 events/kg/year with sense recognition (dot-dashed line).
For the same exposure, the shaded area presents the $3 \sigma$  discovery region.}  
\label{fig:prospect1}
\end{center}
\end{figure}

\subsection{Dark Matter exclusion with Directional detection}
\label{sec:exclu}
At first, one may think of using directional detection to set exclusion limits. Several methods have been proposed 
(\cite{henderson,billard.exclusion}), amongst which the {\em Directional Likelihood exclusion method} (\cite{billard.exclusion}) is the most 
conservative one as it uses only the angular part of the event distribution,  
to avoid assumptions on the unknown energy spectrum of the background. It considers 
the theoretical angular distributions of both WIMP and background events  in order to set the 
most restrictive limits.\\
In the case of a low exposure (30 kg.year) $\rm CF_4$ directional detector, it has been shown that exclusion
limits down to $\sim 10^{-5} \ {\rm pb}$ for highly background-contaminated data or down to $\sim 10^{-6} \ {\rm pb}$ for background-free data
(sensitivity) may be reached, see fig.~\ref{fig:prospect1}. As expected, increasing the
number of background events induces an upward shift of  the   cross
section limit.  However, taking full  advantage on the knowledge of the expected
WIMP and background angular distributions in the data analysis allows to be less sensitive to background contamination. 
This is definitely a major advantage of directional detection.

\subsection{Rejecting the isotropy with directional detection}
Using the clear and unambiguous difference between WIMP signal and background, directional detection may also be used to
  prove that the directional data are not compatible with background. With the help of unbinned likelihood 
  method (\cite{copi}) or non-parametric statistical tests on unbinned data (\cite{green.iso}), it has been shown 
  that a few number of events  ${\cal O} (10)$ is required to reject the isotropy, and hence prove the data are not compatible with the
  expected background. This may give a decisive contribution of directional detection to the field of Dark Matter, especially at the
  present stage when non-directional experiments start to observe candidate events 
  whose origin is difficult to assess (\cite{Ahmed:2011gh,Aprile:2011hi,Aalseth:2010vx,Angloher:2011uu}).

\begin{figure*}[t]
\begin{center} 
\includegraphics[scale=0.4,angle=0]{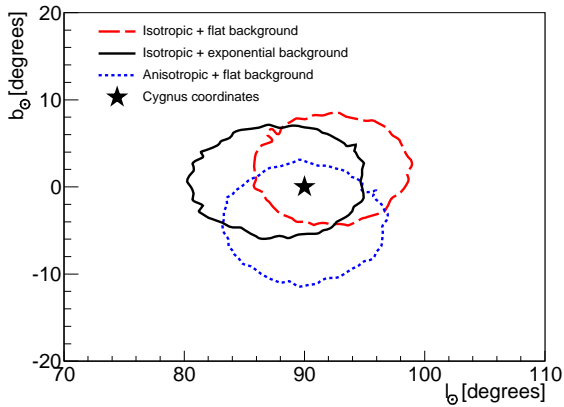}
 \caption{95\% contour level in the ($\ell, b$) plan for various input
models : Isotropic/anisotropic halo model, flat/exponential background. In all cases, the signal is proved to be in the direction of the
Cygnus constellation. Figure from (\cite{billard.ident}).}  
\label{fig:disco}
\end{center}
\end{figure*}

\subsection{Dark Matter discovery with directional detection}
Directional detection may also be used to discover Dark Matter (\cite{billard.disco,green.disco}). In
particular, the method proposed in (\cite{billard.disco})   is a
blind likelihood analysis, the proof of discovery being the fact that the signal points to the direction of 
the Cygnus constellation (to which the solar system's velocity vector is pointing). 
As shown on  fig.~\ref{fig:disco}, the main direction of the incoming events matches the expected direction within  
10$^\circ$ to 20$^\circ$, thus providing a unique signature  of their origin. 
Hence, the goal of this new approach is not to reject
the background hypothesis, but rather to identify a genuine WIMP signal as such.\\
Even at low exposure,  a high significance discovery is achievable, even in the presence of a sizeable background contamination and for
various detector configurations (\cite{Billard.signif}).
Figure \ref{fig:prospect1} presents the region in the ($m_\chi, \sigma_p^{SD}$) plane for which a discovery with a significance greater 
than $3 \sigma$ may be reached with a 30 kg.year $\rm CF_4$ directional detector. It corresponds to a rather large region in the 
SUSY parameter space (\cite{Vasquez:2011bh}), well below current limits from both proton and neutron-based detectors. This highlights the fact that
directional detection is of major interest to clearly identify a positive Dark Matter
detection.

\subsection{Dark Matter identification with directional detection}
\begin{figure*}[t]
\begin{center} 
\includegraphics[scale=0.4,angle=0]{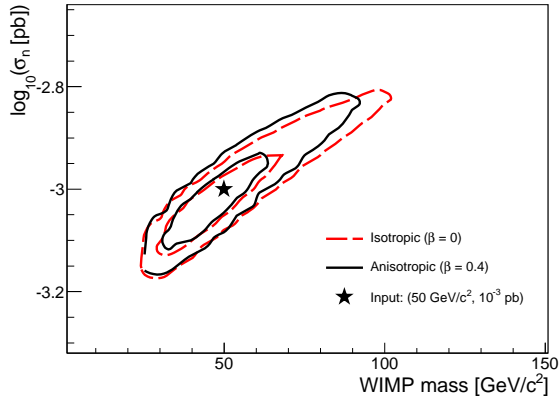}
 \caption{68\% and 95\% contour level in the ($m_{\chi},\sigma_n$) plane, for a 50 $\rm GeV/c^2$ WIMP and 
  two input halo models : isotropic   and 
triaxial. Figure from (\cite{billard.ident}).}  
\label{fig:HaloT4}
\end{center}
\end{figure*}

For high WIMP-nucleon  cross section, it is also possible to go further (\cite{billard.ident}).  
With the help of a high dimensional mutivariate analysis, it is possible  to identify WIMP Dark Matter 
with directional detection. It has been shown that dedicated analysis of 
simulated data of a 30 kg.year $\rm CF_4$ directional detector would allow us to constrain  
the  WIMP properties, both from particle physics ($m_\chi, \sigma_p^{SD}$) and 
galactic halo (velocity dispersions).\\ 
As an example, fig. \ref{fig:HaloT4} presents  the 68\%  and 95\%  contour level in the ($m_{\chi},\sigma_n$) plane. 
This is indeed a measurement of the WIMP properties, consistent
with the input values, with a rather small dispersion and model-independent as the velocity dispersions are set as free
parameters within the framework  of a multivariate Gaussian velocity distribution.

\section{Conclusion}
A 30 kg.year $CF_4$ directional detector would offer a unique opportunity in Dark Matter search, by leading, depending on the value of the unknown  WIMP-nucleon cross section, either to a conclusive exclusion, a high significance discovery 
of galactic Dark Matter or even  an estimation of the WIMP properties.  
However, several key experimental issues need to be addressed to achieve these physical goals, both on the detector side    
and on the data analysis one.


\end{document}